\documentclass[fleqn,usenatbib,useAMS]{mnras}
\usepackage{graphicx}
\usepackage{amsmath}
\usepackage{amssymb}
\usepackage{multicol} 
\usepackage{bm}	
\usepackage{pdflscape}
\usepackage{pdfpages}
\usepackage{caption}
\usepackage{floatrow}
\usepackage{paralist}
\usepackage{verbatim}
\usepackage{caption}
\usepackage{subcaption}
\usepackage{float}
\usepackage{rotating}
\usepackage{times}
\usepackage[T1]{fontenc}
\usepackage{ae,aecompl}
\usepackage{newtxtext,newtxmath}
\usepackage{kantlipsum}
\usepackage{natbib}
\defcitealias{deNicola22_smart_depro}{\scshape Paper II}

\newcommand{\lmod}{\cal{L}_\mathrm{mod}}
\newcommand{\ldat}{\cal{L}_\mathrm{data}}

\newcommand{\aic}{\mathrm{AIC}_\mathrm{p}}

\title[Triaxial Mass Models]{Accuracy and precision of triaxial orbit models I: SMBH mass, stellar mass and dark-matter halo}

\author[B. Neureiter]{B. Neureiter$^{1}$\thanks{E-mail: \href{mailto:mn@ras.org.uk}{bneu@mpe.mpg.de}}
\\
$^{1}$Royal Astronomical Society, Burlington House, Piccadilly, London W1J 0BQ, UK}

\author[B. Neureiter et al.]{
B. Neureiter$^{1,2}$\thanks{E-mail:\href{mailto:bneu@mpe.mpg.de}{bneu@mpe.mpg.de}},
S. de Nicola$^{1,2}$,
J. Thomas$^{1,2}$,
R. Saglia$^{1,2}$,
R. Bender$^{1,2}$
and A. Rantala$^{3}$
\\
$^{1}$Max-Planck-Institut f\"ur Extraterrestriche Physik, Giessenbach-Str. 1, D-85748, Garching, Germany\\
$^{2}$Universit\"ats-Sternwarte M\"unchen, Scheinerstrasse 1, D-81679 M\"unchen, Germany \\
$^{3}$Max-Planck-Institut f\"ur Astrophysik, Karl-Schwarzschild-Str. 1, D-85748, Garching, Germany }
\date{Last updated XXX; in original form YYY}
\pubyear{2020}

\begin{document}
\label{firstpage}
\pagerange{\pageref{firstpage}--\pageref{lastpage}}
\maketitle

\begin{abstract}
We investigate the accuracy and precision of triaxial dynamical orbit models by fitting two dimensional mock observations of a realistic $N$-body merger simulation resembling a massive early-type galaxy with a supermassive black hole (SMBH). 
We show that we can reproduce the triaxial $N$-body merger remnant's correct black hole mass, stellar mass-to-light ratio and total enclosed mass (inside the half-light radius) for several different tested orientations with an unprecedented accuracy of 5-10\%.
Our dynamical models use the entire non-parametric line-of-sight velocity distribution (LOSVD) rather than parametric LOSVDs or velocity moments as constraints. Our results strongly suggest that state-of-the-art integral-field projected kinematic data contain only minor degeneracies with respect to the mass and anisotropy recovery. Moroever, this also demonstrates the strength of the Schwarzschild method in general. We achieve the proven high recovery accuracy and precision with our newly developed modeling machinery by combining several advancements: (i) our new semi-parametric deprojection code probes degeneracies and allows to constrain the viewing angles of a triaxial galaxy; (ii) our new orbit modeling code \texttt{SMART} uses a 5-dim orbital starting space to representatively sample in particular near-Keplerian orbits in galaxy centers;  (iii) we use a generalised information criterion $\aic$ to optimise the smoothing and to compare different mass models to avoid biases that occur in $\chi^2$-based models with varying model flexibilities. 
\end{abstract}


\begingroup
\let\clearpage\relax
\endgroup
\newpage

\section{Introduction}
Early-type galaxies (ETGs) at the high-mass end (absolute magnitude $M_B < -20.5$ mag) bring along particular interesting aspects. They provide information about advanced stages of galaxy evolution, they host the most massive black holes (BHs) observed so far \citep{Mehrgan19}, form in mergers and typically show a central flat core with a tangentially anisotropic orbit distribution (\citealt{Faber97,Bender88_a,Bender89,Gebhardt03,Kormendy09,Gebhardt11,Kormendy13,Thomas14}).
The most massive ETGs also reveal particularities concerning supermassive black hole scaling relations and stellar population analysis.
Growth models for supermassive black holes (SMBHs) present different predictions for the level of scatter at the high-mass end of SMBH scaling relations (\citealt{Peng07,Hirschmann10,Somerville15,Naab17}).
Moreover, it is highly debated whether the stellar initial mass function (IMF) is universal across galaxies or not. Massive ETGs may show the highest fraction of low-mass dwarf stars compared to the Milky Way-like Kroupa- \citep{Kroupa01} or Chabrier- \citep{Chabrier03} IMF (\citealt{Treu10,vanDokkum10,Thomas11,Cappellari12,Spiniello12,Ferreras13,LaBarbera13,Vazdekis15,Smith15,Lyubenova16,Parikh18}). \\
Studying these internal structure and mass composition properties of ETGs is indispensable for understanding massive galaxy formation and evolution. This emphasizes the importance of accurate dynamical modeling routines being able to provide precise information about the intrinsic dynamical structure of ETGs. 
\\
Modeling massive ETGs, however, poses challenges, since specific observational phenomena of massive ETGs, e.g. isophotal twists (\citealt{Bertola79,Williams79,Binney78}), minor axis rotation (\citealt{Schechter78,Contopoulos56,Binney85}) and kinematically decoupled components (\citealt{Bender88_b,Franx88,Statler91,Ene18}) point to a triaxial intrinsic shape of ETGs. Within the SDSS Data Release 3, the bright and massive elliptical galaxies with a de Vaucouleurs profile \citep{deVaucouleurs48} were reported to have a general distribution of the triaxiality parameter\footnote{The triaxiality parameter is defined as $T=\frac{1-p^2}{1-q^2}$ with $q=\frac{c}{a}$ and $p=\frac{b}{a}$, where $a$, $b$ and $c$ are the semi major, intermediate and minor axes of the galaxy.}\citet{Franx91} of 0.4 < T < 0.8 \citep{Vincent05}. Recently, \citet{deNicola22a} used a newly developed semi-parametric triaxial deprojection code \citep{deNicola20} to measure radially resolved shape profiles of individual brightest cluster galaxies. These galaxies are almost maximally triaxial at all radii, however tend to be rounder at their centers compared to their outskirts. \\
Besides the triaxial nature of the stellar components of massive ETGs, most cosmological simulations with collisionless dark matter (DM) halos also predict triaxial DM halo shapes (e.g. \citealt{Jing02,Bailin05,Allgood06,Bett07,Hayashi07,Schneider12,Despali13,Ferrero17}).  \\
The given three dimensional shape of ETGs complicates the extraction of information about their intrinsic properties, since observations only provide a two dimensional projection onto the plane of the sky. The state-of-the-art method to tackle this problem is a dynamical modeling method based on Schwarzschild's orbit superposition technique \citep{Schwarzschild79}. However, it is still unclear how accurate dynamical models in general -- and Schwarzschild models in particular -- can get. The literature about dynamical models addresses a large variety of possible degeneracy issues. For example, \citet{Gerhard96} proved that even in the axisymmetric limit, the deprojection of density distributions is not unique. 
Dynamical models are moreover affected by the well-known mass-anisotropy degeneracy (e.g. \citealt{Gerhard93}), where missing mass in the outer parts, for example, can be hidden by a more tangential orbit distribution.
In the axisymmetric limit, the determination of the correct viewing angles with dynamical modeling routines holds repeatedly stated degeneracies (e.g. \citealt{Krajnovic05,Cappellari06,Onken07,Thomas07b}). Such problems generally increase when going from two to three dimensional systems and the recovery of the orientation and shape of triaxial galaxies has also been reported to be difficult \citep{vandenBosch08}.
Moreover, \citet{Jin19} report large possible stellar and dark-matter mass uncertainties due to the potential degeneracy between them, when analysing galaxies from the Illustris \citep{Vogelsberger14} simulation with a triaxial Schwarzschild modeling routine \citep{vandenBosch08}. \\
Nevertheless, the discussion of scientifically interesting issues, like the previously mentioned questions concerning the stellar IMF and SMBH growth models \textit{demand} correct and accurate black hole mass and stellar mass-to-light ratio recoveries. Fortunately, the last years have provided a lot of progress in various aspects of dynamical modeling such that it is worth to readdress the above degeneracy issues. For example, in the early days of Schwarzschild modeling it was standard to parameterise line-of-sight velocity distributions with Gauss-Hermite moments. Today, it is possible to routinely use the entire, non-parametric line-of-sight velocity distribution \citep{Mehrgan19}, \citet{Barroso21}). The so increased amount of information available certainly helps to overcome some of the degeneracies in earlier models. \\
Also, until recently, the most common method for determining the best fit parameters of dynamical models was a minimization of the observed and modelled discrepancies in a least square sense \citep[e.g.][]{Richstone84, Rix97, Cretton99, Siopis00, Haefner00, Gebhardt00, Valluri04, Thomas04, vandenBosch08, Vasiliev20, Neureiter21}.
However, \citet{Lipka21} showed that the quality of fit between different models cannot be compared with each other without considering the individual model's degrees of freedom. Minimizing a $\chi^2$ across models with varying degrees of freedom leads to biased results. \citet{Thomas22} derived a generalisation of the classical Akaike Information Criterion (AIC) which can be applied to penalised maximum-likelihood models such as most implementations of the Schwarzschild method are. This generalised $\aic$ allows to rigorously include the varying model flexibilities in the comparison of different mass models. Moreover, it allows a data-driven optimisation of the regularisation for each individual trial model \citep{Thomas22}. \\
As another improvement, in our newly developed three dimensional triaxial Schwarzschild Modeling code called \texttt{SMART} \citep{Neureiter21} we use a five-dimensional starting space for orbits to guarantee that all the different orbit types, in particular near the central black hole, are included in the model. \\
Finally, the new semi-parametric deprojection method \texttt{SHAPE3D} introduced by \citet{deNicola20} has shown that the goodness of fit strongly depends on the chosen viewing angles, thus allowing to select the light densities yielding the best $\mathrm{rms}$ \citep[cf.][]{deNicola22a}. \\
All these advancements can potentially reduce the amount of degeneracy in dynamical modeling, and in Schwarzschild modeling in particular.\\
In order to test the combined power and precision of the new
semi-parametric deprojection code \texttt{SHAPE3D} by \citet{deNicola20}, the dynamical modeling routine \texttt{SMART} by \citet{Neureiter21} and the advanced model selection tools developed by \citet{Lipka21} and \citet{Thomas22}, we apply them to high-resolution $N$-body simulations including SMBHs by \citet{Rantala18}. This provides us with the knowledge of the intrinsic scatter and remaining degeneracy uncertainties that one has to deal with when applying triaxial deprojection and dynamical modeling routines to future observational data. In the current paper we focus on the mass reconstruction while in a companion paper by~\citealt{deNicola22_smart_depro},
hereafter called \citetalias{deNicola22_smart_depro}, we discuss the shape and anisotropy recovery.
\\

This paper will be structured as follows: Section~\ref{sec:Triaxial deprojection} and~\ref{sec:Triaxial Schwarzschild code SMART} briefly summarize the used deprojection and dynamical modeling codes. In Section~\ref{sec:The N-Body Simulation} we describe the used $N$-body simulation and our methodology to process its data and model it. In Section~\ref{sec:Results} we present our results, which are then discussed in Sections~\ref{sec: Discussion} and summarized in Section~\ref{sec: Summary and Conclusion}.

\section{Triaxial deprojection}
\label{sec:Triaxial deprojection}
\citet{deNicola20} presented a new semi-parametric deprojection code called \texttt{SHAPE3D} as triaxial extension of the non-parametric axisymmetric algorithm by \citet{Magorrian99}. Triaxial deprojections are highly degenerate. Therefore, one aims for a deprojection method being able to consider all possible density distributions leading to the same projected surface brightness and afterwards evaluate their individual likelihood. Parametric methods like the well known and widely used Multi Gaussian Expansion Method (MGE, introduced by \citet{Monnet92}) are fast, however fail to suggest more than one out of many possible solutions per viewing angle and to select the best light densities using an $\mathrm{rms}$-cutoff. \\
\texttt{SHAPE3D} is, in contrast, able to deal with the degeneracy issue and allows to search for a range of possible deprojections per viewing angle. It is a semi-parametric constrained-shape approach in the sense that it searches for best-fit light densities assuming that the contours of the luminosity density can be described as ellipsoids with possible boxy or discy deformations as well as radially varying axis ratios. 
Under this assumption the galaxy's three dimensional density function $\rho(x,y,z)$ can at every point be described by an ellipsoid whose radius is given as 
\begin{equation}
    m^{2-\xi(x)} = x^{2-\xi(x)} + \left[\frac{y}{p(x)}\right]^{2-\xi(x)}
    							+ \left[\frac{z}{q(x)}\right]^{2-\xi(x)}.
\end{equation}
The four one dimensional functions $\rho(x), p(x), q(x)$ and $\xi(x)$ describe the density, axis ratios $p \equiv y/x$ and $q \equiv z/x$ and the discy- ($\xi > 0$) or boxiness ($\xi < 0$) along the major axis.
The code utilizes a grid-based approach, where the observed surface brightness and density are evaluated on elliptical and ellipsoidal polar grids, respectively. 
Due to the used semi-parametric method, a regularizing penalty function $P$ is necessary to discard unsmooth, non-physical solutions. In total, the code minimizes $L = -\frac{\chi^2}{2} + P$, where $\chi^2$ describes the difference between the observed and modeled SB. 
\citet{deNicola20} proved that their code is able to recover the triaxial intrinsic density of an $N$-body simulation \citep{Rantala18} with high precision when the viewing angles are known. 

Very important for the dynamical modeling is the fact that in the observationally realistic case of unknown viewing directions the assumption of a pseudo-ellipsoidal density structure constrains the range of possible orientations quite strongly \citep{deNicola20}. Moreover, the code filters out deprojections leading to unrealistic $p$- and $q$-profiles, i.e. deprojections which are either not smooth, or are outside the observed shape distribution of massive ellipticals. Furthermore, the code identifies deprojections where the order of the principal axes (short, intermediate, long) changes with radius. Finally, the range of possible viewing directions is narrowed down even more by re-projecting the remaining densities and by evaluating the likelihood of the corresponding isophotal shapes in comparison with the distribution of observed isophotal shapes of ETGs in general (see also \citealt{Thomas05}).

\section{Triaxial Schwarzschild code \texttt{SMART}}
\label{sec:Triaxial Schwarzschild code SMART}
\verb'SMART' is the abbreviation for "Structure and MAss Recovery of Triaxial galaxies" and is a three dimensional implementation of Schwarzschild's Orbit Superposition Technique based on its axisymmetric predecessor by \citet{Thomas04}. We refer to our paper by \citet{Neureiter21} for a detailed description and will only briefly summarize the most important aspects here. 
\begin{enumerate}
 \item \verb'SMART' assembles the total gravitational potential 
\begin{equation}
\Phi=\Phi_{*}+\Phi_{\mathrm{DM}}+\Phi_{\mathrm{SMBH}}
\end{equation}
out of its three relevant contributions. $\Phi_{*}$ and $\Phi_{\mathrm{DM}}$ are the potentials of stars and dark matter (DM). They are computed from the stellar and DM densities (see Section~\ref{sec:Modeling}) via expansion into spherical harmonics. This enables the capability to deal with non-parametric densities. $\Phi_{\mathrm{SMBH}}$ corresponds to the point-like potential from the central supermassive black hole.  
\item \verb'SMART' launches thousands of orbits from a five dimensional starting space and integrates their trajectories for 100 surfaces of section crossings. The five dimensional starting space enables to deal with radially changing structures in the integrals-of-motion-space and therefore allows an automatic adaption to changes in the gravitational potential including a more spherical shape of the potential in the close vicinity of the SMBH giving rise to nearly-Keplerian or rosette orbits (e.g. \citealt{Neureiter21,Frigo21}). 
\item \verb'SMART' fits the kinematic data by computing 
\begin{equation}
    \begin{aligned} \chi^{2}=& \sum_{j}^{N_{\text {losvd }}} \sum_{k}^{N_{\text {vlos }}}\left(\frac{{\ldat}^{j k}-{\lmod}^{j k}}{\Delta {\ldat}^{j k}}\right)^{2}\end{aligned}
    \label{chi squared}
\end{equation}
as the discrepancy between the non-parametric, full LOSVD of the model $\lmod$ and the data $\ldat$ summed over all spatial bins $j$ and velocity bins $k$. $\lmod$ is the sum over the individual orbital LOSVDs weighted by the orbits' occupation numbers, hereafter called orbital weights $w_i$. Since the number of orbital weights as free parameters in general is larger than the total number of observed data consisting as the number of kinematic bins $N_{\text {vlos }}$ times the number of spatial bins $N_{\text {losvd}}$, solving for the orbital weights is underconstrained and the solution ambiguous.
This issue asks for the inclusion of a penalty function.

\item \verb'SMART' therefore conducts the orbit superposition by maximizing and entropy-like quantity 
\begin{equation}
\label{eq:entropy-like quantity}
\hat{S} \equiv S - \alpha \, \chi^2,
\end{equation}
where $\alpha$ is a regularization parameter and
\begin{equation}
\label{eq:entropy term}
S = - \sum_i w_i 
\ln \left( \frac{w_i}{\omega_i} \right).
\end{equation}
The parameters ${\omega_i}$ can be interpreted as weights of the orbital weights $w_i$. The orbital weights $w_i$ are constrained to reproduce the observed photometry as a boundary condition and the specific choice of ${\omega_i}$ defines the chosen entropy term which gets maximized. In our fiducial set-up, we use ${\omega_i}=\mathrm{const}$ so that $S$ is linked to the Shannon entropy. By picking a specific set of ${\omega_i}$ the solution for the orbital weights becomes unique and \texttt{SMART} recovers this solution in the extremely high-dimensional space of the orbital weights with very high precision \citep{Neureiter21}. Different sets of ${\omega_i}$ lead to formally different solutions. We showed in \citet{Neureiter21} that varying the ${\omega_i}$ allows to probe the entire space of possible solutions. However, in the same paper we showed that this modeling freedom does not significantly affect the macroscopic properties of interest such as the mass or anisotropy recovery. Hence, we do not need to explore this additional model space and only use the set $\omega_i = \mathrm{const}$ as described above.

\item In contrast to the orbital weights, the specific choice of the regularization parameter $\alpha$ however does show a notable impact on the model results and achieved precision. It has been shown for axisymmetric models by \citet{Lipka21} that using the optimal smoothing is important to obtain unbiased results. 
To optimise the smoothing in each individual mass model we compute models for a range of different smoothing values\footnote{We typically use $N_\alpha=30$ trial smoothing values distributed homogeneously between $\log \alpha = -6$ and $\log \alpha = 1$.} and select the best one using the generalised information criterion 
\begin{equation}
    \aic = \chi^2+2m_{\mathrm{eff}}
    \label{eq:aic_p}
\end{equation}
for penalised maximum-likelihood models \citep{Thomas22}. In $\aic$ the model flexibility -- which decreases with increasing smoothing strength (i.e., $\alpha \to 0$) -- is represented by the number of effective free parameters $m_\mathrm{eff}$ \citep{Lipka21}. As discussed in more detail in this paper, $m_\mathrm{eff}$ is computed by creating $N_\mathrm{b}$ bootstrap iterations for the LOSVDs, hereafter called $\mathcal{L}_\mathrm{bdata}$, by adding random Gaussian noise based on the observational error $\Delta {\ldat}$ to the original modelled fit $\mathcal{L}_\mathrm{mod}$.
The number of free parameters is then given as as:
\begin{equation}
  m_\mathrm{eff}=\frac{1}{N_\mathrm{b}}\sum_{\mathrm{b}}^{N_\mathrm{b}}  \sum_{n}^{N_{\text {data }}} \frac{1}{(\Delta \mathcal{L}^{n}_\mathrm{data})^2} (\mathcal{L}_\mathrm{bfit}^{\mathrm{b} n} - \mathcal{L}_\mathrm{mod}^{n})(\mathcal{L}_\mathrm{bdata}^{\mathrm{b} n}-\mathcal{L}_\mathrm{mod}^{n}),
\end{equation}
where $N_\mathrm{data}=N_\mathrm{losvd} \times N_\mathrm{vlos}$ is the total number of data points and $\mathcal{L}_\mathrm{bfit}$ is the new modelled fit to the bootstrap data set $\mathcal{L}_\mathrm{bdata}$.\\
It was shown in \citet{Thomas22} that the optimal smoothing is achieved at the minimum of $\aic$. As discussed in detail in the same paper, the smoothing optimisation can be done with a very low number of bootstrap iterations for $m_\mathrm{eff}$. We use $N_\mathrm{b}=1$. We note that the optimal smoothing strength usually varies from model to model. In our case, the closer the assumed mass distribution and orientation are to the true properties of the $N$-body projection, the stronger the optimal smoothing becomes.

\item When evaluating different mass models (or orientations) against each other, the intrinsic model flexibilities vary as described above. The fit qualities cannot be compared to each other in an unbiased manner without taking into account the individual number of the models' degrees of freedom \citep{Lipka21}. Again, we select the best model based on $\aic$. However, here the correlations between different models are weaker than in the case of the smoothing optimisation. As a result, one is often faced with jagged $\chi^2$ curves and also with an increased scatter in $m_\mathrm{eff}$ (cf. the extended discussion in \citealt{Thomas22}). When comparing models obtained with different orbit libraries (i.e. models with different mass distributions and/or with different assumed orientations/shapes) we therefore use $N_\mathrm{b}=15$ bootstrap iterations to calculate an improved estimate of $m_\mathrm{eff}$ at the optimal value of the regularisation parameter $\alpha$ of the individual mass model. As we will show below, with this newly integrated approach we avoid any bias and achieve significantly improved constraints when searching for our best-fit parameters.

\end{enumerate}

\section{The $N$-Body Simulation}
\label{sec:The N-Body Simulation}

We apply our deprojection routine and \verb'SMART' to the high-resolution $N$-body simulation by \citet{Rantala18}.  \\
The simulation is in particular suitable for our application under study since it represents a realistic triaxial remnant of a single generation binary galaxy merger with a structure and shape resembling the core galaxy NGC1600 (e.g. \citealt{Thomas09,Rantala19}). It has a final SMBH of $1.7 \times 10^{10} M_{\odot}$, a sphere of influence\footnote{We here define the sphere of influence as the radius within which the total stellar mass equals the black hole mass.} of $r_{\mathrm{SOI}} \sim1 \mathrm{\,kpc}$ and an effective radius of $r_e\sim14\mathrm{\,kpc}$. \\
The simulation was chosen on purpose for our requested analysis because of its ability to accurately compute the dynamics close to the SMBH due to an algorithmic chain regularization routine AR-CHAIN (\citealt{Mikkola06,Mikkola08}) included in the Gadget-3 (\citealt{Springel05}) based KETJU simulation code (\citealt{Rantala17}).
The used snapshot, which is about one Gyr after the merger has happened, shows a large core and a prolate shape in the outskirts with a more spherical shape towards the center. The stellar component of the merger remnant is maximally triaxial (i.e. $T=0.5$) at $\sim 3$kpc.\\

\subsection{Tested viewing directions}
\label{sec:Tested viewing directions}
We analyse four different projections of this $N$-body simulation: two principal axes of the chosen snapshot as lines of sight ("interm", "minor") as well as one projection exactly in between the principal axes ("middle") and one projection with randomly sampled viewing angles ("rand"). The specific projections and their corresponding viewing angles can be read from table~\ref{tab:projections}. The viewing angles $\theta$ and $\phi$ determine the projection to the plane on the sky and $\psi$ determines the rotation in the plane of the sky. The viewing angles transform the intrinsic coordinates ($x$, $y$, $z$), which are adapted to the symmetry of the object, to the sky-projected coordinates ($x^{\prime}$, $y^{\prime}$, $z^{\prime}$) via the two matrices $P$ and $R$:

\begin{equation}
\left(\begin{array}{l}{x^{\prime}} \\ {y^{\prime}} \\ {z^{\prime}}\end{array}\right)=R \cdot P \cdot\left(\begin{array}{l}{x} \\ {y} \\ {z}\end{array}\right),
\end{equation}
with
\begin{equation}
\label{eq:R matrix}
R =\left(\begin{array}{ccc}{\sin \psi} & {-\cos \psi} & {0} \\ {\cos \psi} & {\sin \psi} & {0} \\ {0} & {0} & {1}\end{array}\right)
\end{equation}
and
\begin{equation}
P =\left(\begin{array}{ccc}{-\sin \phi} & {\cos \phi} & {0} \\ {-\cos \theta \cos \phi} & {-\cos \theta \sin \phi} & {\sin \theta} \\ {\sin \theta \cos \phi} & {\sin \theta \sin \phi} & {\cos \theta}\end{array}\right).
\end{equation}

\begin{table}[h!]
\centering
 \caption{Tested projections with corresponding viewing angles. In order to deproject and model different sets of input data we evaluate the SB and kinematic data of the $N$-body simulation along four different lines of sight. The tested projections include two principal axes, i.e. the intermediate (hereafter called "interm") and "minor" axes of the triaxial merger remnant, as lines of sight. We test another projection exactly in between the principal axes, which we hereafter call "middle", and one more projection (hereafter called "rand") with randomly drawn viewing angles.}
  \label{tab:projections}
 \begin{tabular}{c|c|}
 \hline
 \hline
 \textbf{projection:} & ($\theta,\phi,\psi$) \\
  \hline
 interm &  (90, 90, 90)$^\circ$ \\
 minor &   (0, 90, 90)$^\circ$ \\
 middle &  (45, 45, 45)$^\circ$ \\
 rand &  (60.4, 162.3, 7.5)$^\circ$ \\
  \hline
 \end{tabular}
\end{table}

\subsection{Processing the simulation data}
\label{sec:Processing the simulation data} 
We align the coordinate system of the remnant galaxy to the center of mass and principle axes of the reduced inertia tensor for stars and dark matter within 30 kpc. \\ 
Afterwards, we individually compute the surface brightness (SB) and kinematics for the four different projections under study (see table~\ref{tab:projections}). We assume the galaxy to be in a distance of 20 Mpc.\\
The SB in units of stellar simulation particles is computed with a resolution of 0.1 arcsec within a FOV of ($40 \times 40$) arcsec and a resolution of 0.5 arcsec within ($10300 \times 10300$) arcsec (spanning about 30 times the effective radius). \\
The kinematic data is computed by using the Voronoi tessellation method of \citet{Cappellari03}. 
Our chosen field of view of $\sim 15$ kpc spans about the effective radius. The central Voronoi tesselation within the sphere of influence samples a higher resolution than the tesselation scheme in the outskirts. We compute the Voronoi tesselation grid individually for every projection to guarantee a constant number of stellar simulation particles $N_*$ in each bin. The size of the Voronoi bins is chosen so that the signal-to-noise ratio is 
\begin{align}
    \frac{\mathrm{signal}}{\mathrm{noise}}=\sqrt{\mathrm{signal}}=\sqrt{N_{*}}= \begin{cases} 70 \textrm{ for } r<r_{\mathrm{SOI}} ,\\ 150 \textrm{ for } r_{\mathrm{SOI}}<r<15\mathrm{\,kpc} .  \end{cases}
\end{align}
Averaged over the four tested projections, our kinematic data exhibit $N_\mathrm{voronoi}=220$ Voronoi bins within the whole FOV and $N_\mathrm{voronoi}=51$ Voronoi bins within $r_\mathrm{SOI}$. 
The innermost Voronoi bin spans an average radius of 0.58 arcsec. With this, our chosen resolution matches realistic observational data. \\
For each Voronoi bin we compute the simulation's LOSVDs for $N_{\text {vlos }}= 15$ spanning $v^{\mathrm{max}}_{\mathrm{min}} = \pm 1669 \mathrm{\,km\,s^{-1}}$. 
\\
To provide realistic conditions, which are comparable to future observational data, we add Gaussian noise to the intrinsically noiseless kinematic data of the simulation. We set the standard deviation for the Gaussian scattering as $3\%$ of the maximum of each LOSVD. With this, the velocity dispersion of the noisy kinematic simulation data results in an observationally realistic error of $\sim 2\%$.\\

We have chosen this simulation and mock data setup on purpose, since its high resolution meets the requirements of our study. 
 Our goal is to demonstrate the accuracy and precision that can be achieved with advanced dynamical models and the best current observational data. The actual precision in any specific measurement will depend on the circumstances, e.g. signal-to-noise ratio in the spectral observations, spatial resolution, distance and many other factors. It is not an intrinsic property of the modeling process. However, better data do not necessarily guarantee better results. In particular for dynamical modelling, the existence of intrinsic degeneracies (e.g. between mass and anisotropy) may eventually limit the achievable precision regardless of the quality of the data. However, while often discussed, the effect of such degeneracies has rarely been quantified. Our goal here is to show that they do not hamper highly accurate dynamical measurements on a 10\% level.

\subsection{Modeling the N-body simulation}
\label{sec:Modeling}
We first apply the deprojection routine to the different tested projections of the $N$-body simulation as if dealing with an observed galaxy. For each tested projection, the original grid of 1800 trial viewing angles is reduced by the deprojection code to a few dozen candidate orientations or shapes, respectively (see also \citetalias{deNicola22_smart_depro}).\\
Similar to when modeling real observational data, we model a multidimensional parameter space and do not provide any a priori knowledge about the analysed $N$-body merger. 
Besides the viewing angles $\theta$, $\phi$ and $\psi$, we vary the black hole mass $M_{BH}$, the stellar mass-to-light ratio $\Upsilon$ as well as five dark matter halo parameters. The DM halo profile is parameterised similar to a generalized NFW model with a scale radius $r_s$, density normalisation $\rho_0$, axis ratios $p_\mathrm{DM}$ and $q_\mathrm{DM}$ and a variable inner logarithmic density slope $\mathrm{\gamma_{in}}$. However, because the original halos of the $N$-body progenitor galaxies were based on a Hernquist profile, we set the outer logarithmic density slope equal to  $\mathrm{\gamma_{out}}=-4.5$ (measured value for the remnant of our merger simulation) rather than the canonical NFW value of $\mathrm{\gamma_{out}}=-3.0$. \\
The individual parameters of our ten dimensional parameter space are each sampled on a grid and the 
best-fit parameters are determined by looking for the minimum in $\mathrm{AIC}_p$ (see Equation~\ref{eq:aic_p} in Section~\ref{sec:Triaxial Schwarzschild code SMART}). We do this individually for each tested projection. The results for the best-fit viewing angles are detailed in \citetalias{deNicola22_smart_depro}. There, it is shown that the viewing angles are recovered with an average deviation of $\sim 20^\circ$.\\
In the current paper we focus on the detailed recovery of the stellar mass-to-light ratio and black hole mass. 
The sampled grid for the stellar mass-to-light ratio covers 10 values within $\Upsilon \in [0.6,1.4]$. The corresponding grid size is $\Delta \Upsilon = 0.09$, which equals 9\% of the true value $\Upsilon_\mathrm{sim}=1$ of the simulation. The grid for the black hole mass covers 10 values within $M_{BH} \in [1.0,3.0] \times 10^{10} M_{\odot}$ ($\Delta M_{BH} = 0.22 \times 10^{10} M_\odot$, i.e. 13\% of the true black hole mass $M_{BH,\mathrm{sim}}=1.7 \times 10^{10} M_{\odot}$). \\
The model results presented in this paper are attained by modeling each projection
twice: 
by taking advantage of the triaxial symmetry of the simulation we can split the kinematic data of each projection along, e.g., the apparent minor axis (determined by averaging over the projected isophotes). One data set shows only positive values for the sky-projected coordinates $x'>0$ (hereafter called 'right side' of the galaxy) and the other data set shows only negative values for the long axis of the sky-projection coordinates, i.e. $x'<0$ (hereafter called 'left side' of the galaxy). This provides us with two independent kinematic data-sets for each tested projection, allowing us to determine the scatter of our modeling results.
While it is actually common practice in Schwarzschild models to derive parameter uncertainties from a $\Delta \chi^2$-criterion, we specifically chose to use a different method. 
Since \citet{Lipka21} showed that the effective number of parameters in the models varies with mass, viewing angles, etc. and that the Schwarzschild fitting is a model selection process rather than a parameter optimisation, a $\Delta \chi^2$-criterion is statistically meaningless. An easy and unbiased way to determine errors in such a situation is to use the actual scatter ‘measured' over fits to several data sets (see~\citealt{Lipka21}). \\
As our results in Section~\ref{sec:Results} will demonstrate, this method results in robust estimates of the actual scatter in the best-fit parameters. \\

For every kinematic data set we ran on average 3000 models to sample our ten dimensional parameter space. We use the software NOMAD (Nonlinear Optimisation by Mesh Adaptive Direct search; \citealt{Audet06,LeDigabel11}) to optimise the search. NOMAD is designed for time-consuming constrained so-called black-box optimisation problems. \\

\section{Results}
\label{sec:Results}
In \citetalias{deNicola22_smart_depro}, we discuss the recovery of the intrinsic shape, including the recovery of the viewing angles $\theta, \phi, \psi$, axis ratios $p,q$ and triaxiality parameter $T$, as well as anisotropy $\beta$ of the $N$-body simulation, while we here focus on the question how well we can recover the mass distribution. \\

\begin{figure*}
    \centering
    \includegraphics[width=1.0\textwidth]{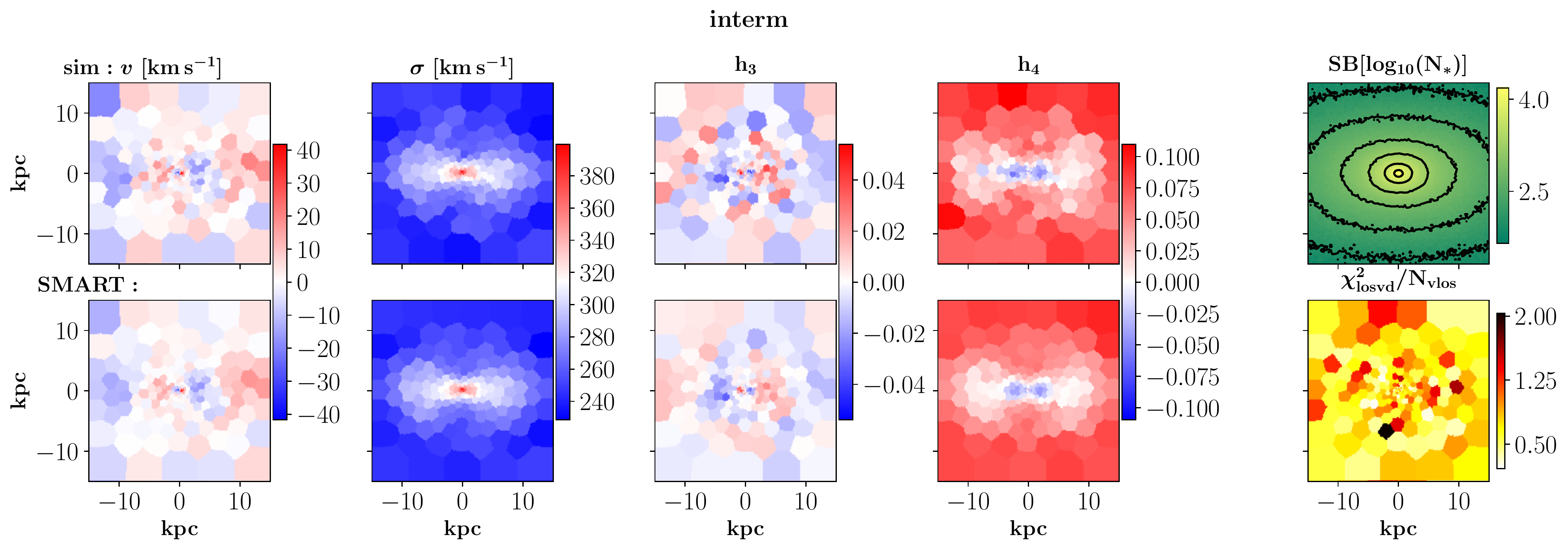}
    \caption{Velocity (first column), velocity dispersion (second column), $h_3$ (third column) and $h_4$ (fourth column) map of the simulation (first row) and the best-fit model (second row) for the intermediate axis projection ('right' data set). The true viewing angles are $\theta=\phi=\psi=90^\circ$ in this case, the recovered ones are $\theta=60^\circ, \phi=90^\circ, \psi=90^\circ$ (see \citetalias{deNicola22_smart_depro}). \texttt{SMART} is able to accurately fit the kinematic input data with $\chi^2/\mathrm{N_{data}}=0.62$. The $\chi^2/\mathrm{N_{vlos}}$-map (bottom right panel) shows that the fit can be produced homogeneously well over the whole field of view. The top right panel shows the surface brightness map in logarithmic units of stellar particle numbers $N_*$ of the simulation.}
    \label{fig:velmap_interm}
\end{figure*}

\texttt{SMART} fits the kinematic input data very well, independent of the chosen projection. The average goodness-of-fit is $\Delta \chi^2/\mathrm{N_{data}} = 0.69$, where $N_\mathrm{data}=N_\mathrm{vlos}\times N_\mathrm{voronoi}$ consists of the number of velocity bins $N_\mathrm{vlos}$ times the number of Voronoi bins $N_\mathrm{voronoi}$ of the individual projection and respective modeled side.
Fig.~\ref{fig:velmap_interm} shows maps of the velocity, velocity dispersion, as well as the Gauss-Hermite parameters $h_3$ and $h_4$ \citep{Gerhard93,vanderMarel93,Bender94}. The Figure shows both the input data and the model fit and we have chosen the intermediate-axis projection as an example. The velocity maps for the minor, middle and rand projections are plotted in Apendix~\ref{sec: Velocity maps}. As one can see, the modeled maps match the input kinematic data homogeneously well over the whole field of view.  In particular, \texttt{SMART} is able to reproduce the negative $h_4$-parameter in the center, which corresponds to a tangentially anisotropic orbit distribution produced during the core formation process. Note, that we do not fit the Gauss-Hermite moments but instead fit the entire non-parametric LOSVDs at $N_\mathrm{vlos}$ line-of-sight velocities $v_\mathrm{los}$ in each Voronoi bin. We show the  Gauss-Hermite maps only to illustrate the fit quality. However, the Figure also shows the goodness-of-fit $\Delta \chi^2/\mathrm{N_{vlos}}$ achieved over the entire non-parametric LOSVD in each individual Voronoi bin. \\

\begin{figure*}[h!]
    \centering
    \includegraphics[width=0.8\textwidth]{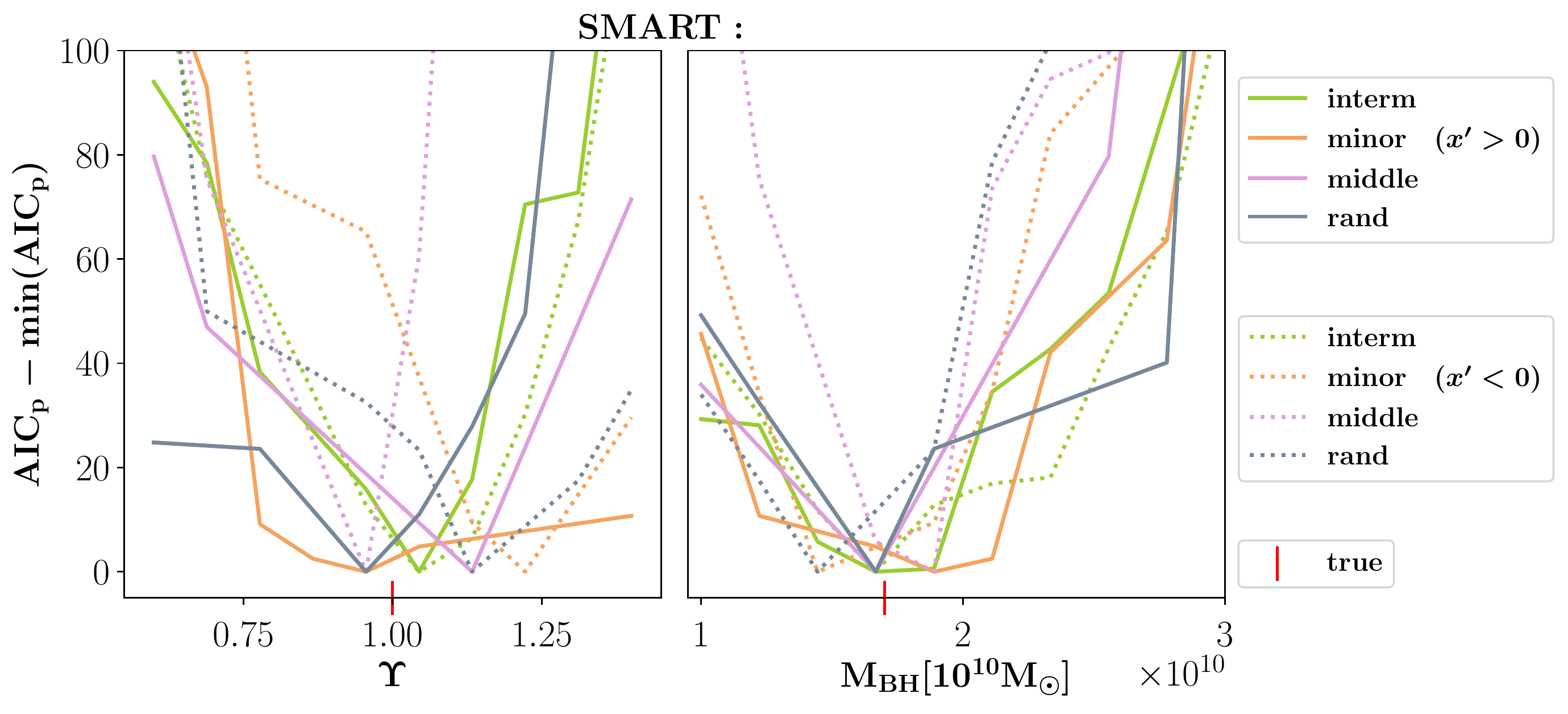}
    \caption{Mass recovery results of $\Upsilon$ (left panel) and $M_{BH}$ (right panel) for the four tested projections (different colors) and their respective modeled sides (solid lines for the right sides and dotted lines for the left sides). The minima of the $\aic$-curves point at the best-fit stellar mass-to-light ratio (left panel) and black hole mass (right panel) of the individual models, which cover a multi-dimensional parameter space. The true mass values from the simulation ($\Upsilon_\mathrm{sim} = 1 $ and $M_{BH,\mathrm{sim}}= 1.7 \times 10^{10} M_\odot$) are marked with red lines. For each modelled projection, the best-fit values consistently scatter within 10\% of the 
    true values (see Table~\ref{tab:individual recovery}). With this, we are able to reproduce the correct mass parameters with an unprecedented accuracy. This achieved precision indicates that, in general, projected kinematic data of a triaxial galaxy contain only minor degeneracies.}
    \label{fig:ml_mbh_triax_results}
\end{figure*}

Fig.~\ref{fig:ml_mbh_triax_results} shows the curves of $\aic$ versus the tested stellar mass-to-light ratios and black hole masses for the different projections and sides. For these curves we first search at each $\Upsilon$ (or  $M_\mathrm{BH}$, respectively) the minimum $\aic$ over all other parameters and then connect these values. As already mentioned in Section~\ref{sec:Modeling}, the final best-fit model is determined as the global minimum of all $\aic$ values over all parameters, hereafter called min($\aic$). 
The absolute $\aic$-values of the various projections/sides are not important. They cannot be compared to each other, because every data set has a different number of kinematic input data (see section~\ref{sec:Processing the simulation data}). We therefore subtract the individual min($\aic$) from the respective $\aic$ values of the same data set. Each $\aic$ curve represents the results of $\sim 3000$ models in the ten dimensional space of the mass and orientation parameters. \\

As one can see, all best-fit stellar mass-to-light ratios and black hole masses scatter within a small variation range around the true values of the simulation (red lines). 
For future studies it is interesting to investigate the accuracy and precision of individual measurements for observational data with similar resolution and coverage as assumed in this study. For this we analyse the mean black hole masses and stellar mass-to-light ratios and their corresponding standard deviations for the two sides of each individual projections (as in detail explained in Section \ref{sec:Modeling}).
The results for the individual measurements are summarized in table~\ref{tab:individual recovery}. Within the individual standard deviations, the black hole mass as well as the stellar mass-to-light ratio were correctly recovered on the $10\%$ accuracy level. As one can see, with our choice of averaging over two independent data-sets, we yield representative scatter measurements, which are in the same order of magnitude for each tested projection.

\begin{table}[h!]
\centering
 \caption{Recovery precision of $\Upsilon$ and $M_{BH}$ for individual measurements. In order to estimate the precision level one can expect when analysing future observational data with a resolution similar to the one from the N-body simulation in the current analysis, we here individually list the results of the black hole mass and stellar mass-to-light ratio of the four tested projections and compare them to the true values from the simulation. Within the standard deviations, which are given by modeling the two sides of each projection, every tested data-set correctly recovers the true values from the simulation with a minor deviation on the $\sim 5 - 10\%$ level.}
  \label{tab:individual recovery}
 \begin{tabular}{c|c|c}
 &  $\Upsilon$ & $M_{BH}$  \\
  \hline
 interm & $ 1.04 \pm 0.05$ & $(1.67 \pm 0.07)\times 10^{10} M_\odot$ \\
 minor & $ 1.09 \pm 0.13$ & $(1.67 \pm 0.22)\times 10^{10} M_\odot$   \\
 middle &    $ 1.05 \pm 0.08$ & $(1.78 \pm 0.11)\times 10^{10} M_\odot$\\
 rand &  $ 1.05 \pm 0.08$ & $(1.56 \pm 0.11)\times 10^{10} M_\odot$  \\
 \hline \hline
 true & $\Upsilon_\mathrm{sim}=1$ & $M_{BH,\mathrm{sim}} = 1.7 \times 10^{10} M_\odot$\\
 \end{tabular}
\end{table}

In addition to the precision of individual measurements it is also important to study the statistical accuracy, which can in principle be achieved with an accurate triaxial dynamical modeling machinery. Due to the fact that we analyse several mock samples by modeling different projections, we can determine an average accuracy of the method.
Averaged over the results of the interm, minor, middle and rand projection, we achieve $\Upsilon = 1.06 \pm 0.09 $ and $M_\mathrm{BH}= (1.67 \pm 0.16)\times 10^{10} M_\odot$.  With this, the mean stellar mass-to-light ratio is recovered within $\Delta \Upsilon= 6 \%$ and the mean black hole mass is recovered within $\Delta M_\mathrm{BH}= 2 \%$ in comparison to the true values ($\Upsilon_{sim}=1$, $M_{BH,\mathrm{sim}}=1.7 \times 10^{10} M_{\odot}$) of the simulation. The true values lie within the standard deviation of our tested models. 
This accuracy is slightly below our considered grid step sizes of $\Delta_\mathrm{grid} \Upsilon = \pm 9\%$ and $\Delta_\mathrm{grid} M_\mathrm{BH} = \pm 13\%$. We therefore estimate conservatively that the accuracy is at least the grid step size, i.e. of the order of 10\%, though it is probably even better. \\

In order to provide a complete test in our analysis along all principal axes of a triaxial galaxy, we also performed models along the long axis of the analysed $N$-body galaxy and found that the discussed results change for this particular line of sight. 
For our fiducial ten dimensional parameter space setup, the best-fit black hole masses derived from the major-axis projection of the $N$-body are off by more than $70 \%$. In Appendix~\ref{sec: major axis analysis} we show that these offsets vanish when we assume the right orientation and radial shape of the DM halo profile, i.e. when we provide the normalized stellar and DM halo density profiles of the simulation and/or when we increase the input data resolution. Hence, these offsets are not related to \texttt{SMART}, but instead indicate that the dynamical modeling and in particular the recovery of the exact black hole mass of a triaxial galaxy gets more difficult when a galaxy happens to be observed along its intrinsic long axis. This would not be entirely surprising since we know that even the pseudo-ellipsoidal deprojections become degenerate when an object happens to be observed along one of its principal axes. Kinematic degeneracies are likely largest for viewing-angles along the principal axes as well. However, such viewing angles -- in particular if only the major axis is concerned -- are rare and an increased uncertainty along this direction will not severely affect the results of triaxial models for randomly selected galaxies. Nevertheless, we plan a more in-depth analysis of this particular case and its implications in a future paper.

\begin{figure*}[h!]
    \centering
    \includegraphics[width=0.75\textwidth]{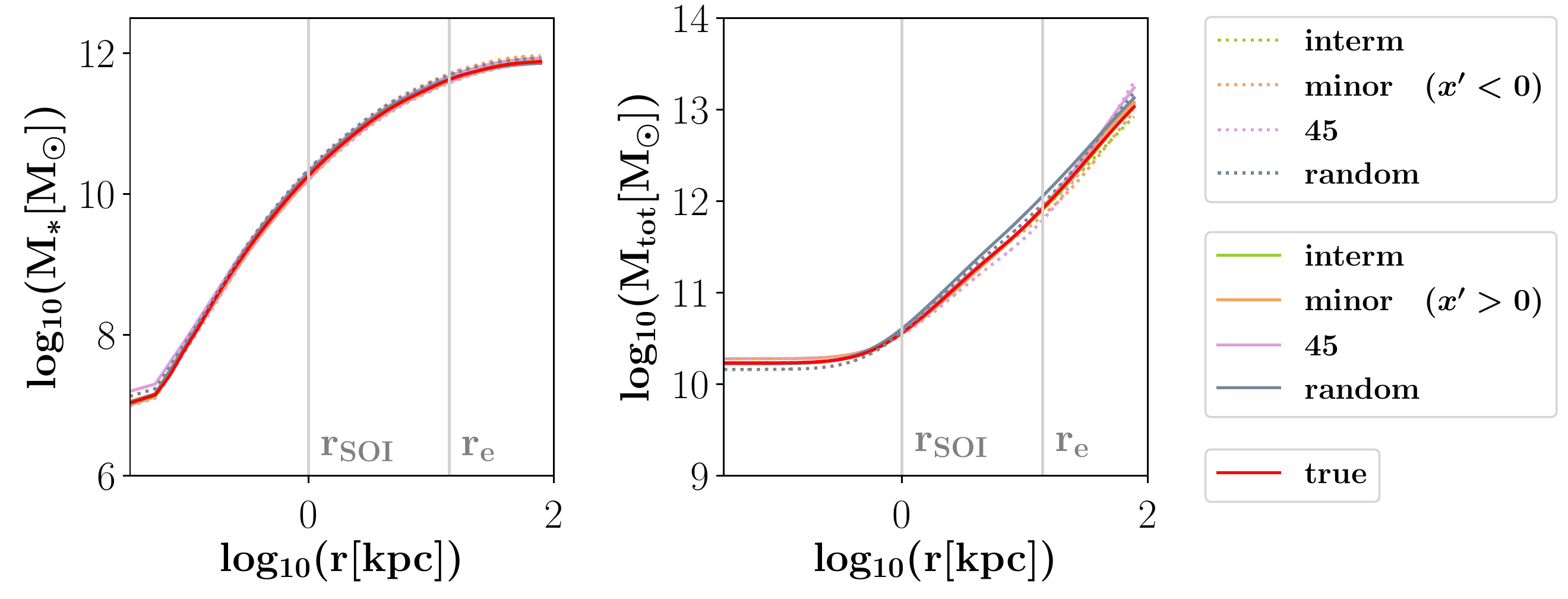}
    \caption{Mass recovery results of the enclosed stellar (left panel) and total mass (right panel) profiles. The simulation's sphere of influence and effective radius are marked with grey lines. For $r<r_\mathrm{SOI}$  the black hole is the dominant mass contributor. The stellar mass dominates within $r_\mathrm{SOI}<r<r_e$ and the DM mass dominates for $r>r_e$. The stellar and total enclosed mass profiles from the different projections and modeled sides (different colors) follow the real one from the simulation (red line) over all radii. The average deviation of the stellar mass is only $\Delta M_\mathrm{*}(7\mathrm{kpc})=5.9\%$ at an intermediate radius of 7kpc and the average deviation of the total enclosed mass is $\Delta M_\mathrm{tot}(r_\mathrm{SOI})=5.9 \%$ at the sphere of influence and $\Delta M_\mathrm{tot}(r_e)=4.5 \%$ at the effective radius.}
    \label{fig:encl_mass_triax_results}
\end{figure*}

Fig.~\ref{fig:encl_mass_triax_results} shows the recovery of the enclosed stellar (left panel) and total mass (right panel) profiles for the interm, minor, middle and rand projection. The total mass consists of the sum of the black hole, stellar and DM mass. 
Within $r_{\mathrm{SOI}} < r < r_e$ the stellar part dominates over the BH and dark matter. At a distance of $r \sim ~r_e$, the enclosed DM mass equals the enclosed stellar mass of the simulation. Therefore, for radii $r>r_e$ the DM mass is the main contribution to the enclosed total mass, whereas the BH mass dominates the mass contribution within the sphere of influence. 
As one can see, the stellar enclosed mass profiles of all best-fit models (different colors) follow the real one from the simulation (red) over all radii, in particular within the relevant radial region between $r_{\mathrm{SOI}} < r < r_e$, and even down to a radius, where the stellar mass is less than 10\% of $M_\mathrm{BH}$. At an intermediate radius of 7kpc the mean deviation from the stellar enclosed mass between the best-fit models and the simulation is only $\Delta M_\mathrm{*}(7\mathrm{kpc})=5.9\%$. Also the total enclosed mass profiles of all best-fit models follow the real one from the simulation over all radii. It follows that also the enclosed DM profile is well reproduced. 
Averaged over the four different projections, the relative deviation from the total enclosed mass is $\Delta M_\mathrm{tot}(r_\mathrm{SOI})=5.9 \%$ at the sphere of influence and $\Delta M_\mathrm{tot}(r_e)=4.5 \%$ at the effective radius.\\
Besides the accurate recovery of the total enclosed mass, \texttt{SMART} is furthermore able to determine the correct, non-spherical shape of the DM halo. Averaged over our tested projections and sides, the axis ratios $p_\mathrm{DM}=0.79 \pm 0.06$ and $q_\mathrm{DM}=0.91 \pm 0.06$ show a maximum deviation of the true values $p_\mathrm{DM,sim}=q_\mathrm{DM,sim}=0.93$ of only 0.14. 
The principal axis ratios $p_\mathrm{DM,sim}, q_\mathrm{DM,sim}$ are thereby computed via the eigenvalues from the reduced inertia tensor of the simulated DM particles within 100 kpc.
Our findings show that our triaxial deprojection and orbit modeling codes prove to produce reliable mass recovery results for the SMBH, stellar and DM components with an accuracy on the $\sim 5 - 10\%$ level. \\
Of course the precision of individual measurements depends on specific circumstances like the signal-to-noise ratio of the data, their spatial resolution etc. Hence the above numbers do not imply that every measurement will have this precision. However, our tests demonstrate that a $5-10\%$ level of precision is possible with appropriate data and advanced Schwarzschild models.\\ 
In \citetalias{deNicola22_smart_depro} we show that a similar level of accuracy is achieved for the orbital anisotropy. This provides a rigorous test for our modeling machinery.

All these results together strongly suggest that, in principle, the intrinsic degeneracies contained in the photometric data and in particular in state-of-the-art integral field kinematic data are small enough so that macroscopic parameters of interest like the mass of the central SMBH, the stellar mass-to-light ratio and also the anisotropy profile (cf. \citetalias{deNicola22_smart_depro}) of a triaxial galaxy can be determined with better than 10\% precision. In this sense, this sets a reference for the astonishing small \textit{intrinsic scatter} of triaxial dynamical modeling routines, which can be expected and achieved for precise kinematic data comparable to the N-body's resolution. \\ Special caution must be paid, when a triaxial galaxy is observed along its long axis (App.~\ref{sec: major axis analysis}).

\section{Discussion}
\label{sec: Discussion}

The results presented in the previous Sec.~\ref{sec:Results} suggest that the accuracy and precision that can be achieved with (triaxial) dynamical orbit models is much better than previously anticipated.

\begin{figure*}[h!]
    \centering
    \includegraphics[width=0.7\textwidth]{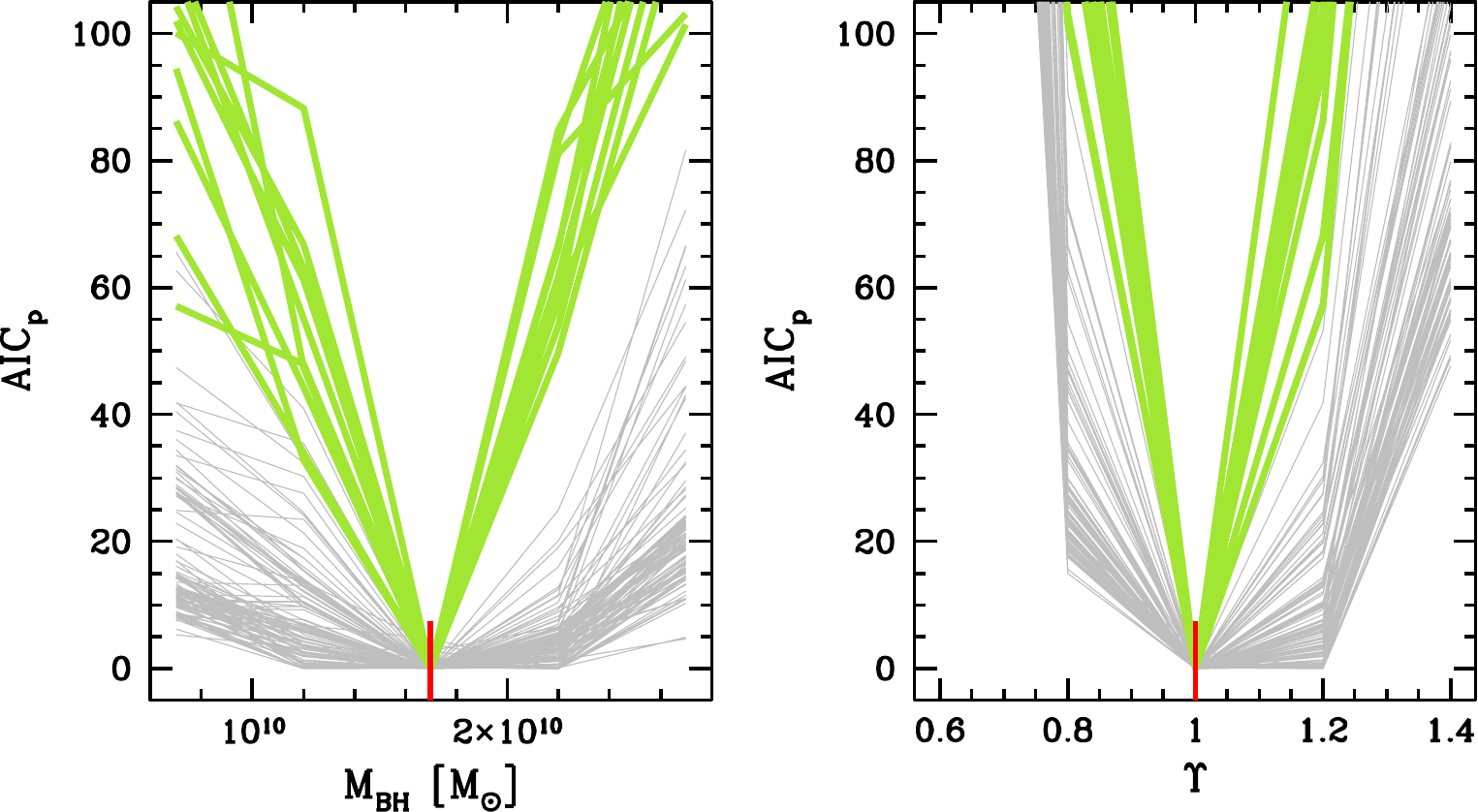}
    \caption{Constraints on the black-hole mass $M_\mathrm{BH}$ (left) and the stellar mass-to-light ratio $\Upsilon$ (right) in idealised model fits. As input data we use ten different mock realisations of the interm projection. For the model fits we assume the true DM halo, the true viewing angles and the true three dimensional stellar light profile -- only $M_\mathrm{BH}$ and the normalisation of the stellar mass are assumed to be unknown. The grey lines show $\chi^2$ curves derived using a constant regularisation parameter $\alpha$. We only show results for $\alpha$ values that lead to acceptable fits ($\chi^2<N_\mathrm{data}$). This does not uniquely determine $\alpha$. Typically, all models for $\alpha > \sim 10^{-3}$ provide such acceptable fits. We show for all ten mocks all of the $\chi^2$ curves resulting from these different assumed plausible smoothings. The green lines show the $\aic$ curves derived as described in Sec.~\ref{sec:Triaxial Schwarzschild code SMART}. The red vertical lines mark the true values of the $N$-body simulation. The figure shows that even under the idealised conditions assumed here (halo, orientation and stellar light profile known), the $\chi^2$ minimisation allows for a wide range of different solutions and that the best-fit $\Upsilon$-values are often strongly biased ($\Upsilon \approx 1.0 - 1.2$). The fact that $\chi^2$-based mass derivations can be biased high was already discussed in \citet{Lipka21} for axisymmetric models. This bias disappears and the constraints \textit{improve significantly} when the model selection via $\aic$ is applied -- for both the optimisation of the smoothing and the comparison of different mass models.}
    \label{fig:model_selection}
\end{figure*}

\subsection{Importance of model selection}\label{sec: Importance of model selection}
We used the same triaxial $N$-body simulation already in \citet{Neureiter21} to show that the central SMBH mass, the stellar mass-to-light ratio and $\beta$-profile can be recovered to better than a few percent accuracy with our dynamical models. In that paper, we assumed the angles and the stellar and DM density shape to be known and focused on testing our orbit modeling code \texttt{SMART}, i.e. did not go through all the analysis steps of a real galaxy.  Here we go one step further. We simulate the entire modeling process of an observed galaxy. The difference to \citet{Neureiter21} is not only that we here use \textit{noisy} input data but we also simulate the realistic situation where we do not know the galaxy's orientation and intrinsic shape because we only have its projected image on the sky. Still, the mass and anisotropy recovery results of our current studies (see also \citetalias{deNicola22_smart_depro}) reach a similar precision ($\sim 5 - 10\%$) as in the idealised case studied in \citet{Neureiter21}. On the one side, this reaffirms our previous tests and suggests that even in the realistic case where one has to deal with (i) noisy data and (ii) a situation where the intrinsic shape and orientation are unknown, an almost unique solution for the macroscopic parameters of interest of a triaxial galaxy can be found. 
However, on the other side it is surprising that even though the number of unknowns in the modeling process has increased so much, we still reach a comparable precision as in \citet{Neureiter21}. A substantial difference between the current work and the work presented in \citet{Neureiter21} is the way in which we choose our best-fit model. 
The results presented in \citet{Neureiter21} were evaluated at values of $\alpha$, for which the internal velocity dispersions of the $N$-body simulation were best recovered by the model. This information is of course not available for real observational data. \\
Therefore, in the current paper we use the approach explained in Section~\ref{sec:Triaxial Schwarzschild code SMART}: We optimise the smoothing for each individual orbit library using a purely data-driven method. This allows the smoothing to adapt to the particular data set at hand and varies from one mass model to the other. \\
To illustrate how important this smoothing optimisation and the $\aic$ comparison (cf. Section~\ref{sec:Triaxial Schwarzschild code SMART}) of different models is, we remodelled ten different mock realisations of the interm projection of the $N$-body simulation using a very idealised model setup: we assumed the DM halo to be known, the three dimensional stellar light profile to be known and the viewing angles to be known. Only $M_\mathrm{BH}$ and $\Upsilon$ were treated as free parameters. In classical Schwarzschild applications the $\chi^2$ (see eq. \ref{chi squared}) would be minimised for some constant value of the smoothing value $\alpha$. In Fig.~\ref{fig:model_selection} we illustrate this case by the grey lines. Each line shows the modeling results of a classical $\chi^2$ minimisation for some constant value of $\alpha$. We consider only smoothing values for which the minimum obtained $\chi^2 < N_\mathrm{data}$ -- i.e. only smoothing values that lead to acceptable best-fit models. As the figure shows, even in this highly idealised case, where almost all properties are known to the model, the optimisation of the remaining two paramters $M_\mathrm{BH}$ and $\Upsilon$ leads to results with unsatisfyingly large uncertainties ($\sim$30\% for $M_\mathrm{BH}$). Moreover, the values for $\Upsilon$ tend to be biased high by up to 20\%. In comparison, the model selection using $\aic$ and adaptive optimised smoothing is much more accurate and precise (see green lines in Fig.~\ref{fig:model_selection}). The fact that the $\chi^2$ minimisation in this case, where almost every property of the model is assumed to be known, results in uncertainties/biases much larger than for our fiducial full modeling shows how important the correct model selection is to reach the accuracy and precision that we reported above. \\

 Another possible way to calibrate the relative strength of goodness-of-fit -- measured by the $\chi^2$ -- and the strength of the smoothing would be through Monte-Carlo simulations. Based on a toy model with known properties one tests different smoothing strengths and checks which one allows for the best recovery. This (constant) smoothing strength is then used for the analysis of observed galaxies. This method is expensive since in principle it should be repeated for each individual data set with its characteristic individual error distributions, spatial coverage etc. and for each galaxy with its characteristic individual orbital structure, shape etc. It is also uncertain since there is no guarantee that the toy model used for calibration has the same structure as the galaxy to be analysed. In this context, we want to stress here, that the optimal smoothing in our case even depends on which projection of the $N$-body we analyse -- even though it is always the \textit{same} $N$-body simulation that we fit. \\
 Since all Schwarzschild codes use some sort of regularisation in order to avoid overfitting (e.g. \citealt{Richstone88,Merritt93,Verolme02,Thomas04,Valluri04,vandenBosch08,Vasiliev20,Neureiter21}), the question of how to choose an optimised regularization becomes crucial when it comes to the level of high accuracy and precision that we could achieve with \texttt{SMART}.

\subsection{Comparison to other triaxial Schwarzschild Models}

\citet{vandenBosch09} modelled 13 simulated photometric and kinematic data resembling SAURON (e.g.~\citealt{Emsellem04}) observations from possible oblate fast rotators to triaxial slow rotators. They skipped any recovery of black hole masses and concentrated on recovering the intrinsic shape and stellar mass-to-light ratio. They correctly recovered $\Upsilon$ within $10\%$ for the cases with well recovered intrinsic shape and within $20\%$ for the cases with less constrained intrinsic shapes. 
Using the same code, \citet{Jin19} modelled 9 triaxial early-type galaxies from the high resolution Illustris simulation. They were able to recover the total enclosed mass within the effective radius with $15\%$ accuracy and an underestimation of the stellar mass of $\sim24\%$. Again, no recovery of black holes was included in the study. A direct comparison remains difficult because different studies assume different input data.
Nevertheless, the unprecedented precision that we achieved in our tests highlights the importance of extensive methodology verifications, e.g. by application to high resolution $N$-body simulations. The application of triaxial dynamical models with an unexpected high precision as demonstrated here for our code to future observational data promises interesting new results from stellar dynamics.

\section{Summary and Conclusion}
\label{sec: Summary and Conclusion}
We have presented the updated version of our modeling machinery and its efficiency by application to an $N$-body merger simulation resembling a realistic massive early-type galaxy hosting a supermassive black hole. In order to create realistic conditions we compute the triaxial merger remnant's kinematic two dimensional data on a Voronoi binning with a spatial resolution comparable to today’s telescopes’ resolution. We furthermore add a plausible amount of Gaussian noise and evaluate the kinematic data with a velocity resolution similar to future observational data. 
Our modeling machinery implements several features:
\begin{enumerate}
    \item[(i)] To provide our dynamical modeling code \texttt{SMART} with a predecision on possible deprojections and viewing angles, we use the flexible new semi-parametric triaxial deprojection code \texttt{SHAPE3D} \citep[cf.][]{deNicola20}.
    \item[(ii)] \texttt{SMART} reconstructs the stellar orbit distribution by integrating thousands of orbits, which are launched from a five dimensional starting space to cover all orbital shapes in particular near the central black hole \citep[cf.][]{Neureiter21}.
    \item[(iii)] \texttt{SMART} exploits the full non-parametrically sampled LOSVDs rather than using velocity moments as constraints \citep[cf.][]{Neureiter21}.
    \item[(iv)]  \texttt{SMART} uses an adaptive smoothing scheme to optimise the regularisation in each trial mass model \citep[cf.][]{Thomas22}.
    \item[(v)] \texttt{SMART} uses a generalised information criterion for penalised models to select the best-fitting orbit model, avoiding potential biases in $\chi^2$-based approaches \citep[cf.][]{Lipka21,Thomas22}.
\end{enumerate}
Similar to the case of observed galaxies, we model a multi-dimensional parameter space, including the a priori unknown viewing angles as well as the mass parameters for the stellar and DM components of the galaxy's potential. In order to test multiple mock samples, we apply our triaxial deprojection code \texttt{SHAPE3D} and modeling code \texttt{SMART} to four different projections from the $N$-body simulation.\\
\texttt{SMART} is able to fit the kinematic input data homogeneously well over the whole field of view with a mean accuracy of $\Delta \chi^2/\mathrm{N_{data}} = 0.69$. \\
For each modelled projection, we are able to reconstruct the true stellar mass-to-light ratio $\Upsilon_\mathrm{sim} = 1 $ and black hole mass $M_{BH,\mathrm{sim}}= 1.7 \times 10^{10} M_\odot$ of the simulation with an acuracy on the $\sim 5-10 \%$ level. \\
Also the enclosed total mass profile was correctly recovered by \texttt{SMART} over all radii. The average deviation of the total enclosed mass, consisting of the black hole, stellar and DM mass contributions, is only $\Delta M_\mathrm{tot}(r_\mathrm{SOI})=5.9 \%$ at the sphere of influence and $\Delta M_\mathrm{tot}(r_e)=4.5 \%$ at the effective radius. \\
We are furthermore able to recover the correct, non-spherical shape of the simulation's DM halo by recovering the true axis ratios $p_\mathrm{DM}$ and $q_\mathrm{DM}$ with a maximum deviation of only 0.14. \\
As more extensively presented in our companion \citetalias{deNicola22_smart_depro} by~\citealt{deNicola22_smart_depro}, we are also able to reconstruct the simulation's shape and anisotropy with similar accuracy. We refer to this paper for an extensive discussion of the recovery results for the viewing angles $\theta, \phi, \psi$, axis ratios $p,q$ and orbital anisotropy.\\
The surprisingly high accuracy and precision as well as low degree of degeneracy that we find in our models reaffirm our earlier results presented in \citet{Neureiter21}. There, in an idealised setting with known viewing angles and known deprojection we found that macroscopic parameters of a triaxial galaxy, like the anisotropy and mass composition, are not severely influenced by any degeneracy remaining in the reconstruction of the orbit distribution function.
We can now go one step further. Our results strongly suggest that in general, the projected kinematic data of a triaxial galaxy hold only minor degeneracies, which enables an unentangled recovery of the intrinsic structure and mass composition. \\
With this analysis we were able to show that the intrinsic scatter of accurate triaxial dynamical modeling routines, which are applied to precise kinematic data, is small enough to target scientific questions concerning the scatter of SMBH scaling relations and the well known IMF issue. \\
Our study points to a possible change of this statement for the analysis of a triaxial galaxy observed along its long axis, which will be more extensively studied in a future paper. \\
Another study covered by a future paper will be the axisymmetric analysis of a triaxial galaxy.

\section*{Acknowledgements}
This research was supported by the Excellence Cluster ORIGINS which is funded by the Deutsche Forschungsgemeinschaft (DFG, German Research Foundation) under Germany's Excellence Strategy - EXC-2094-390783311. We used the computing facilities of the Computational Center for Particle and Astrophysics (C2PAP). Computations were performed on the HPC systems Raven and Cobra at the Max Planck Computing and Data Facility

\section*{Data Availability Statement}
The data underlying this article will be shared on reasonable request to the corresponding author.

\bibliographystyle{mnras}
\bibliography{main}

\appendix

\section{major axis analysis}
\label{sec: major axis analysis}
\begin{figure*}
    \centering
    \includegraphics[width=0.8\textwidth]{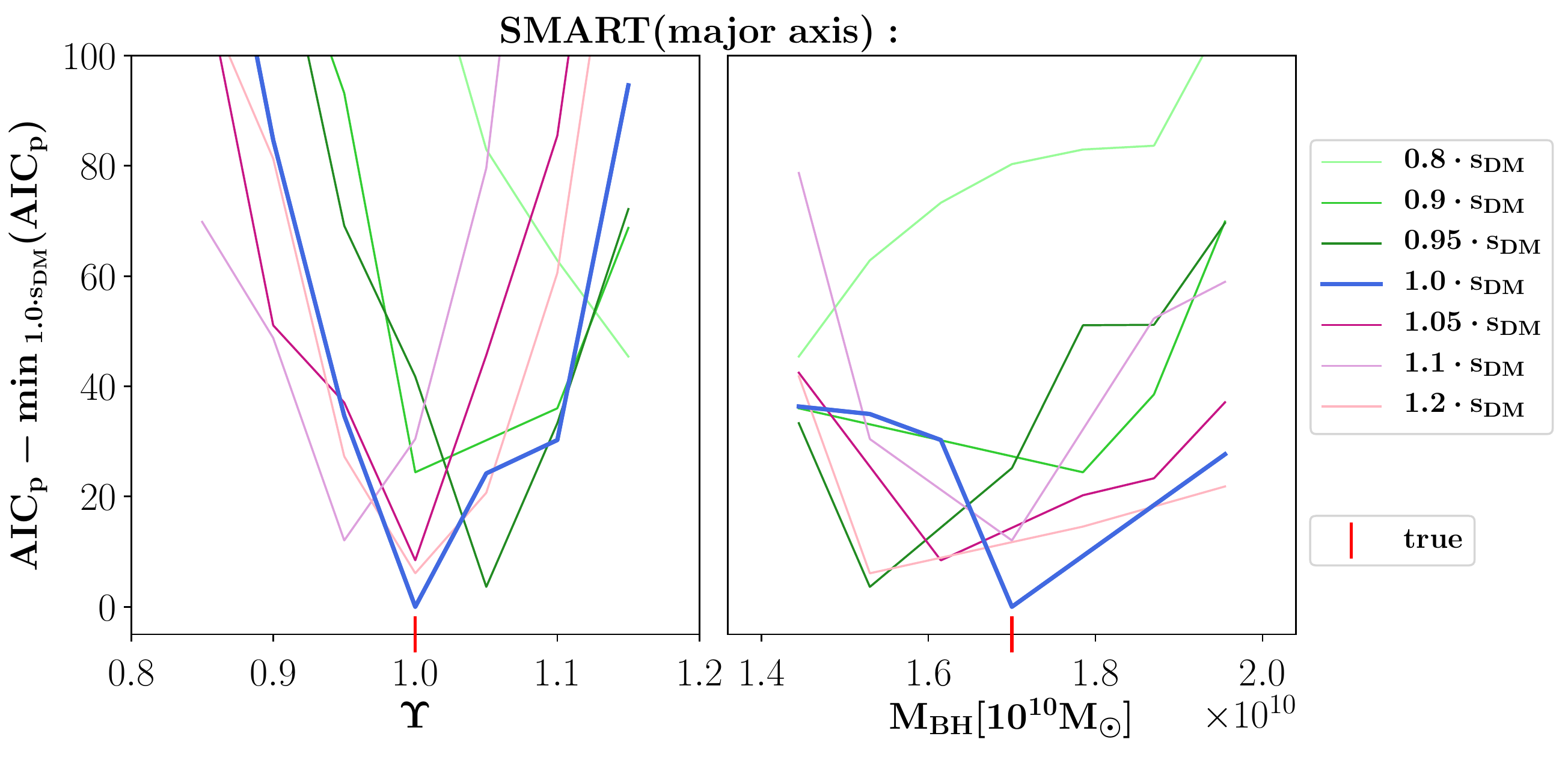}
    \caption{Mass recovery results of $\Upsilon$ (left panel), $M_{BH}$ (right panel) and dark matter scaling factor $s_\mathrm{DM}$ (different colors) for the major-axis projection. In this analysis we remodel a higher resolved kinematic input data of the right side of the major axis by providing \texttt{SMART} with the true normalized stellar and DM density of the simulation. For a better comparison we subtract the minimum $\aic$-value of the models with the correct $s_\mathrm{DM}$-parameter, i.e. $\mathrm{min_{1.0s_{DM}(AIC_p)}}$ (blue line), from all $\aic$ curves. The individual minima for the different colored $\aic$-curves provide the respective best-fit $\Upsilon$- and $M_{BH}$-values for the models with different $s_\mathrm{DM}$-input values. It turns out, that the best-fit model of all tested 343 models covering this three dimensional parameter space, is the one with the true $\Upsilon_\mathrm{sim}$- and $M_{BH,\mathrm{sim}}$-values from the simulation (red) as well as the correct $s_\mathrm{DM}$-parameter (minimum of all curves). This indicates that \texttt{SMART} shows no intrinsic bug when modeling kinematic data with high enough resolution projected along the major axis of a triaxial galaxy.}
    \label{fig:major_axis_analysis}
\end{figure*}

When modeling the ten dimensional parameter space of the $N$-body simulation projected along its major axis, we find different results in comparison to the other tested projections (cf. Section~\ref{sec:Results}).
Along this specific line of sight, the stellar mass-to-light ratio is reproduced with a maximum uncertainty of only 20\% (best-fit stellar mass-to-light ratio for the right side of the major axis projection is $\Upsilon(x^{\prime}>0)=1.04$ and $\Upsilon(x^{\prime}<0)=1.22$ for the left side), however, the best-fit black hole mass is more than 70\% underestimated. \\ 
In order to check that the black hole mass uncertainty along the major axis projection is not caused by an intrinsic bug of \texttt{SMART} along this axis, we remodeled a three dimensional mass parameter grid for the right side of the kinematic data for this axis. 
To minimize modeling uncertainties that originate from incomplete sampling in the ten dimensional parameter space or from uncertainties in the deprojection, we make the following simplifications: 
We do not provide \texttt{SMART} with plausible deprojections determined by the triaxial deprojection routine from \citet{deNicola20}, but we forward the true normalized stellar density from the simulation to \texttt{SMART}. Also, instead of modeling a gNFW halo with five unknown parameters, as used for the analysis in Section~\ref{sec:Results}, we here model the DM halo  by providing \texttt{SMART} with the correct normalized DM density of the simulation with an unknown scaling parameter $s_\mathrm{DM}$. This is the same Ansatz as used in \citet{Neureiter21}. The remaining three parameters for this analysis to be determined by \texttt{SMART} are $\Upsilon$, $M_{BH}$ and $s_\mathrm{DM}$. \\
We also increase the resolution of the kinematic input data from $N_\mathrm{vlos}=15$ (see Section~\ref{sec:Processing the simulation data}) to $N_\mathrm{vlos}=45$. This allows us to fit the kinematic input data with a velocity resolution of $\Delta v_{\mathrm{vlos}} = 71.1 \mathrm{\,km\,s^{-1}}$ instead of the lower velocity resolution of $\Delta v_{\mathrm{vlos}} = 223.5 \mathrm{\,km\,s^{-1}}$ used for the more time-consuming analysis of Section~\ref{sec:Results}.\\ 
For this adapted set-up we evaluate 343 models with different $\Upsilon$-, $M_{BH}$- and  $s_\mathrm{DM}$-input-masses. The tested mass grid covers a grid size of 5\% around the correct mass parameters. \\
Fig.~\ref{fig:major_axis_analysis} shows the outcome of this analysis, where we plot the $\aic$-curves of the models with different  $s_\mathrm{DM}$-input-masses (different colors) against the tested $\Upsilon$- (left panel) and $M_{BH}$- values (right panel). Since all models are provided with the same number of kinematic input data $N_\mathrm{data}$, their absolute $\aic$-values can be compared with each other and their total minimum provides the best-fit model. \texttt{SMART} is able to determine the true stellar mass-to-light ratio, black hole mass as well as DM scaling parameter. \\
This test enables us to show that \texttt{SMART} is in principle able to recover the correct mass parameters for kinematic data projected along the long axis of a triaxial galaxy. \\
The uncertainty of the black hole mass recovery of $70\%$, which was achieved within our fiducial ten dimensional parameter space setup with unknown stellar and DM shape, therefore appears to origin from uncertainties caused by the deprojection and/or the multi-dimensional DM halo modeling and/or the lower resolution of the tested parameter grid size as well as of the kinematic input data, which was used for reasons of computational time. \\
Of course, a non-negligible intrinsic physical degeneracy along this axis cannot be excluded.
A more detailed study of this apparent major-axis abnormality will be performed in the future.

\section{\texorpdfstring{Velocity-, Surface Brightness- and $\chi^2$-maps}{Velocity-, Surface Brightness- and chi-maps}}
\label{sec: Velocity maps}
\begin{figure}
    \centering
    \includegraphics[width=0.9\textwidth]{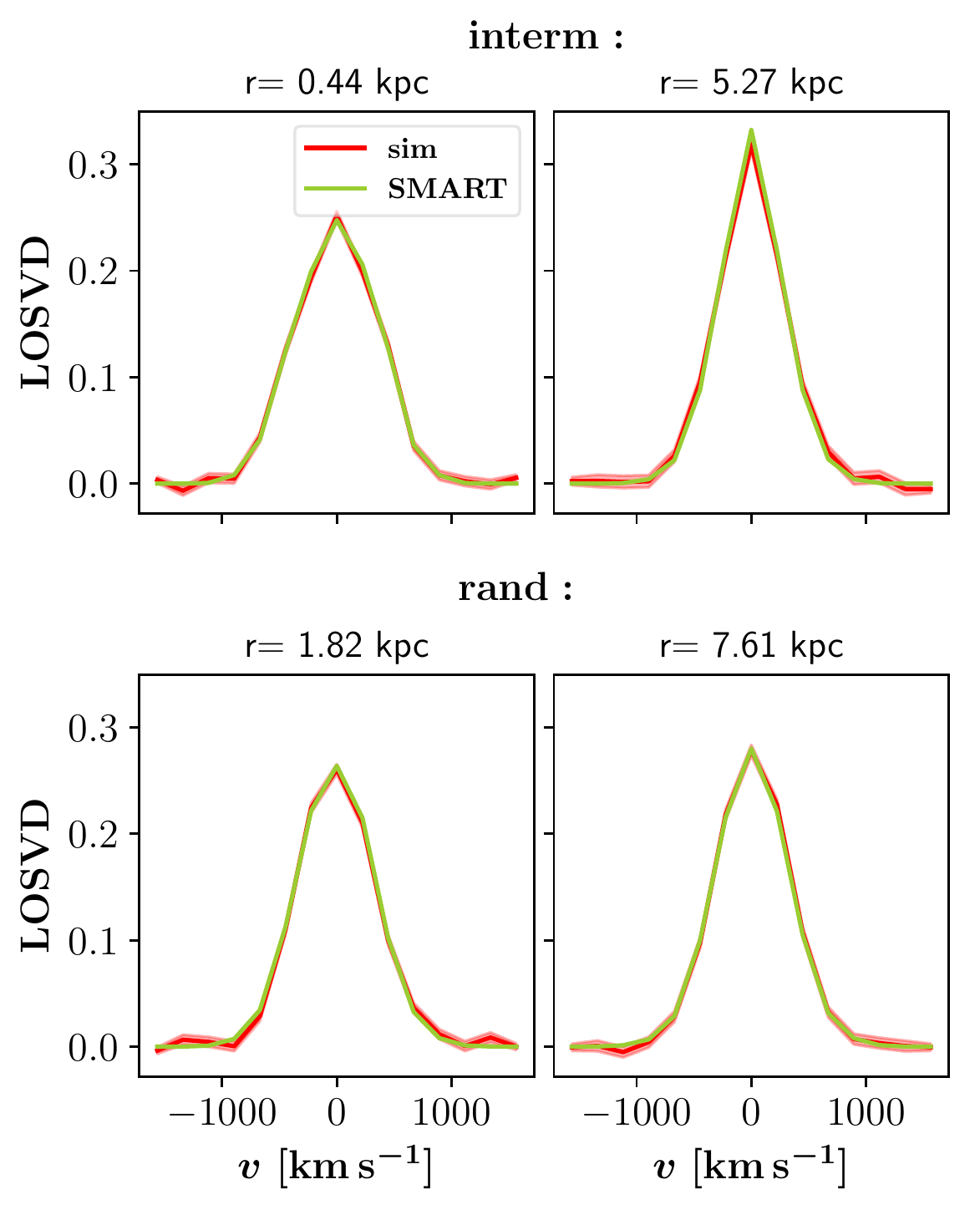}
    \caption{Exemplary demonstration of the achieved LOSVD fit (green lines) in comparison to the input LOSVDs (red lines) for the interm projection (top panels) and for the rand projection (bottom panels) at two different radii, respectively.}
    \label{fig:LOSVD}
\end{figure}

\begin{figure*}
  \centering
\begin{subfigure}[c]{1.0\textwidth}
    \includegraphics[width=1.0\textwidth]{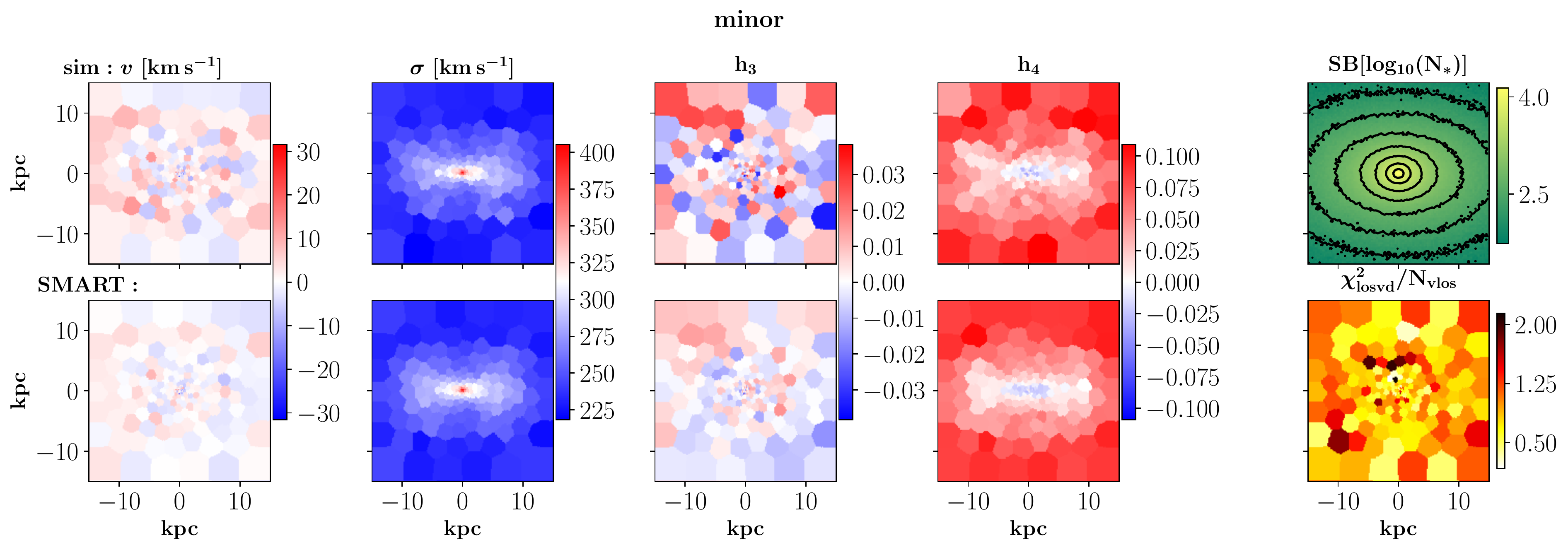} \\[\abovecaptionskip]
  \end{subfigure}
\begin{subfigure}[c]{1.0\textwidth}
    \includegraphics[width=1.0\textwidth]{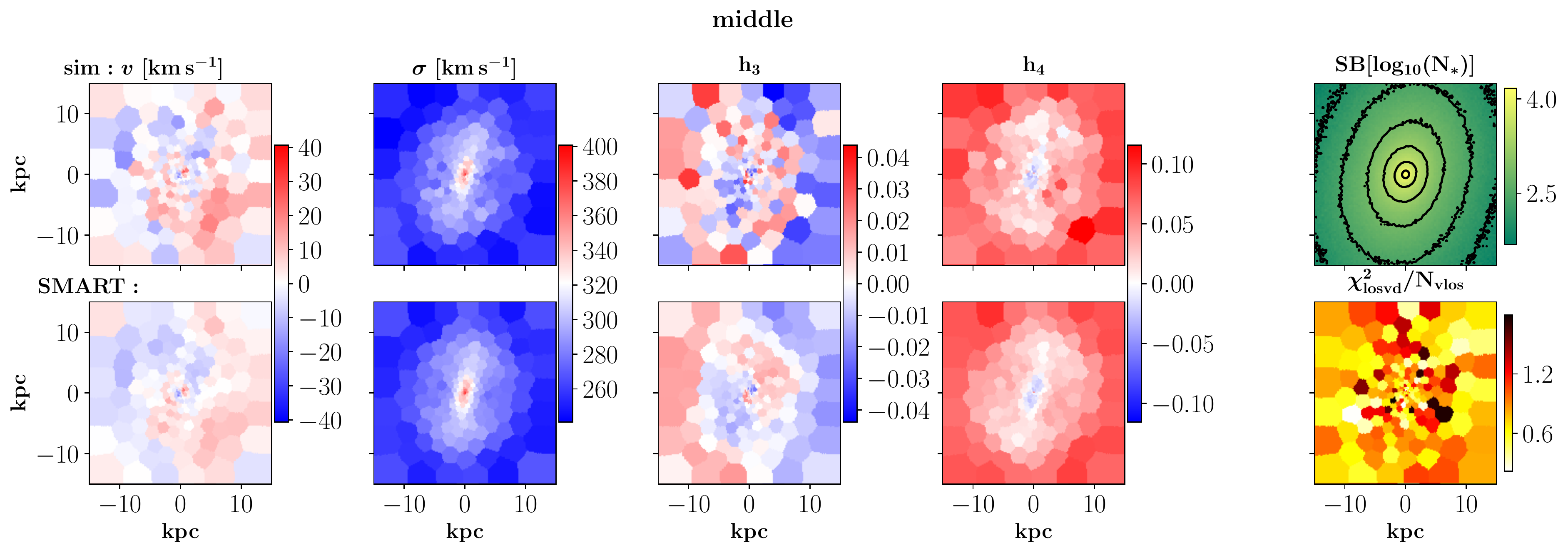} \\[\abovecaptionskip]
  \end{subfigure}
  \begin{subfigure}[c]{1.0\textwidth}
    \includegraphics[width=1.0\textwidth]{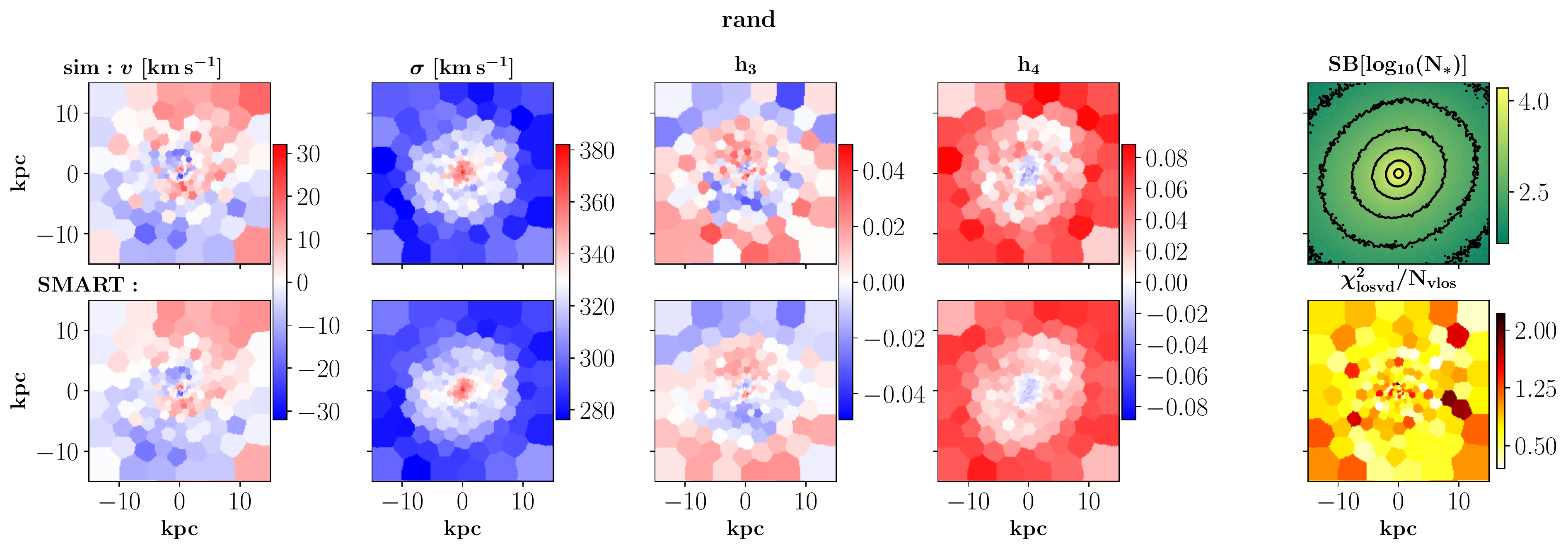} \\[\abovecaptionskip]
  \end{subfigure}
\caption{Velocity, velocity dispersion, $h_3$, $h_4$ and surface brightness maps of the simulation (top row) as well as velocity maps of the best-fit model and corresponding $\chi^2$-map (bottom row) for different projections. The individual line of sight for the different projections can be read from the title. The velocity maps for the intermediate axis were already shown in Fig.~\ref{fig:velmap_interm}. Overall, \texttt{SMART} is able to well fit the kinematic input data independent of the individual viewing angles. The average deviation from the kinematic input data with the modeled fit is $\Delta \chi^2/\mathrm{N_{data}} = 0.69$. The $\chi^2$-maps indicate \texttt{SMART}'s ability to fit the kinematic input data homogeneously well over the whole field of view.}
\label{fig: velmap_minor_45_random}
\end{figure*}

Fig.~\ref{fig: velmap_minor_45_random} shows the velocity $v$, velocity dispersion $\sigma$, $h_3$ and $h_4$ maps of the simulation and kinematic fit for the minor, middle and rand axis as line of sight. In addition, the surface brightness of the simulation is plotted in logarithmic units of stellar simulation particles. The $\chi^2$-map shows the deviation between the kinematic input data and the best-fit model evaluated by \texttt{SMART}. As already stated in Section~\ref{sec:Results}, \texttt{SMART} fits the kinematic input data well for all tested axes over the whole field of view with an average deviation between the input and modelled LOSVDs of $\Delta \chi^2/\mathrm{N_{data}} = 0.69$. The maps of the Gauss-Hermite parameters in Fig.~\ref{fig: velmap_minor_45_random} are only to illustrate the fit quality. \texttt{SMART} actually fits the entire LOSVDs. To demonstrate the fit of the true LOSVD data, Fig.~\ref{fig:LOSVD} shows two input LOSVDs (red lines) and two fitted LOSVDs (green lines) for a central Voronoi bin and an outer Voronoi bin projected along two different lines-of-sight.

\bsp
\label{lastpage}
\end{document}